\documentclass[prl,showpacs,showkeys,twocolumn,floatfix]{revtex4}

\usepackage{graphicx}
\usepackage{units}
\usepackage{xspace}
\usepackage[american]{babel}

\newcommand{\cro}{\ensuremath{{}^{52}}Cr\xspace}
\newcommand{\muB}{\ensuremath{\mu_{B}}}

\begin{document}


\title{Observation of Feshbach resonances in an ultracold gas of ${}^{52}$Cr}

\author{J.\ Werner}
\author{A.\ Griesmaier}
\author{S.\ Hensler}
\author{J.\ Stuhler}
\author{T.\ Pfau}
\email{t.pfau@physik.uni-stuttgart.de}
\affiliation{5.\ Physikalisches Institut, Universit\"at Stuttgart, 70550
Stuttgart, Germany}
\author{A.\ Simoni}
\author{E.\ Tiesinga}
\affiliation{National Institute of Standards and Technology, Gaithersburg,
Maryland 20899-8423, USA}

\date{\today}

\begin{abstract}
  We have observed Feshbach resonances in elastic collisions between
  ultracold \cro atoms. This is the first observation of collisional
  Feshbach resonances in an atomic species with more than one valence
  electron. The zero nuclear spin of \cro and thus the absence of a
  Fermi-contact interaction leads to regularly-spaced resonance
  sequences. By comparing resonance positions with multi-channel
  scattering calculations we determine the $s$-wave scattering length
  of the lowest $^{2S+1}\Sigma_{g}^{+}$ potentials to be
  $\unit[112(14)]{a_0}$, $\unit[58(6)]{a_0}$ and $-\unit[7(20)]{a_0}$
  for $S=6$, 4, and 2, respectively, where $
  a_{0}=\unit[0.0529]{nm}$.
\end{abstract}

\maketitle

With the development of laser cooling and trapping techniques, atomic
collisional properties in the ultracold regime have become directly
accessible. Today, these properties play a crucial role in experiments
with quantum degenerate bosonic and fermionic gases. In the ultracold
regime, elastic collisions between most neutral atoms are dominated by
isotropic interaction potentials, which only depend on the
internuclear separation $R$ and can be characterized by a single
length, the $s$-wave scattering length $a$. This type of interaction
is responsible for many of the fascinating phenomena observed in
Bose-Einstein condensates (BEC's) (for a review see \cite{Bongs:2004})
and degenerate Fermi gases \cite{Chin:2004}.

In alkali-metal gases, the effect of the isotropic potentials and,
consequently, the value of the scattering length can be controlled by
magnetically tunable Feshbach resonances \cite{Tiesinga1993a}.
Feshbach resonances appear when the energy of the incoming atoms
equals the energy of a bound molecular level of a higher-lying
molecular potential and are used to change sign and magnitude of $a$ \cite{Duine:2004}.
Recently, Feshbach resonances have been exploited to study the strong
interaction regime in ultracold atomic gases or even to produce
molecular Bose-Einstein Condensates \cite{Duine:2004}. Feshbach
resonances between different atomic species have also been
theoretically predicted \cite{simoni:163202}, and experimentally
observed \cite{Stan:2004,inouye:183201}.

The spins of the six electrons in the 3d and 4s valence shells of the
${}^{7}$S${}_{3}$ ground state of \cro are aligned.  This gives rise
to a magnetic moment as large as $\mu=\unit[6]{\muB}$, where $\mu_B$ is
the Bohr magneton.  This large magnetic moment is responsible for a
very strong anisotropic spin-spin dipole interaction between two
${}^{7}$S${}_{3}$ \cro atoms.  In fact, when compared to alkali-metal
atoms, which have a maximum magnetic moment of 1 $\mu_B$, it is 36 times
stronger.  This difference has limited hopes of Bose condensing \cro
prepared in a state where the electron spins are aligned parallel to
an external magnetic field by severely restricting its
lifetime \cite{Hensler:2003a}.

For atomic \cro in spin state anti-parallel to the
magnetic field the spin-spin dipole interaction, however, does not lead to atom
losses.  Instead, the effects of the {\it anisotropic} and long-range spin-spin
dipole interaction can add a new twist to the field of ultracold
quantum gases. In particular, the expansion and the stability of a
dipolar BEC is expected to depend on the trapping geometry
\cite{Giovanazzi:2003a}. A roton is expected to develop in the
dispersion relation and new quantum phases in optical lattices have
been predicted for dipolar gases \cite{Baranov:2002b}.  The
anisotropic interaction can be changed by time-varying magnetic fields
\cite{Giovanazzi:2002a}, while the isotropic interaction can be tuned using a
Feshbach resonance. This allows one to arbitrarily adjust the ratio of the isotropic and anisotropic interactions.

Isotropic interactions between two ground-state \cro atoms are due to
Hund's case (a) $^{2S+1}\Sigma^+_{g/u}$ Born-Oppenheimer potentials.
The large number of valence electrons leads to seven Born-Oppenheimer
potentials instead of two for ground-state alkali-metal atoms.
Here, $S$ is the total electron spin of the two atoms and $g/u$ is
{\it gerade}/{\it ungrade} for inversion symmetry of the electron
wavefunction around the center of charge.  For \cro
even (odd) $S$ implies $g$ ($u$) symmetry, respectively.
Conventional spectroscopic data only exists
for the ground-state ${}^{1}\Sigma_{g}^{+}$ potential.
Theoretical {\it ab-initio} calculations
\cite{Andersson:1995,Pavlovic:2004} exist but are
extremely challenging for \cro.

Chromium has been trapped using buffer gas cooling techniques
\cite{Carvalho:2003}, in magneto-optical traps
\cite{McClelland:98,Bell:99}, and in magnetic traps
\cite{Stuhler:2001,Schmidt:2003b}. Using a cross-dimensional relaxation technique,
our group was able to determine the decatriplet $^{13}\Sigma^+_g$ $s$-wave
scattering length of ${}^{52}$Cr to be $\unit[170(39)]{a_{0}}$ and of
${}^{50}$Cr to be $\unit[40(15)]{a_{0}}$ \cite{Schmidt:2003b}. The uncertainty in parenthesis is
a one-standard deviation uncertainty combining statistical and systematic errors.

In this Letter, we report the observation of magnetic Feshbach
resonances in a gas of ultracold \cro atoms. We locate 14
resonances through inelastic loss measurements between magnetic
field values of \unit[0]{mT} and \unit[60]{mT}. The broadest observed
feature has a $1/\sqrt{e}$-width of $\unit[68]{\mu T}$.  By comparing the
experimental data with theoretical multi-channel calculations, we are
able to identify the resonances and to determine the scattering
lengths of the ${}^{13,9,5}\Sigma_{g}^{+}$ Born-Oppenheimer potentials,
the Van der Waals dispersion coefficient $C_6$, and $C_8$, which are
the same for all seven Born-Oppenheimer potentials.

The details of our cooling scheme are presented in
\cite{Schmidt:2003a, Schmidt:2003c}.  After Doppler-cooling in a
clover-leaf type magnetic trap \cite{Schmidt:2003c} and evaporative
cooling we load the atoms into a crossed optical dipole trap. The dipole trap
is realized using an Yb-fiber laser with a wavelength of
\unit[1064]{nm}. The two trapping beams have a waist of
\unit[30]{$\mu$m} and \unit[50]{$\mu$m} and a power of \unit[11]{W} and
\unit[6]{W}, respectively. To suppress dipolar relaxation
\cite{Hensler:2003a}, we optically pump the atoms from the $m_{s}=+3$
Zeeman sublevel of the ${}^{7}S_{3}$ state to the energetically lowest
$m_{s}=-3$ level. To achieve this, we use a frequency-doubled
master-slave diode laser system, which is resonant with the
${}^{7}S_{3}\to{}^{7}P_{3}$ transition at \unit[427.6]{nm}. Using
\unit[250]{$\mu$W} of $\sigma^{-}$ polarized light and an optical pumping
time of \unit[1.2]{ms}, we achieve a transfer efficiency close to 100\%.
The lifetime in the optical dipole trap increases from \unit[7]{s} in
the $m_{s}=+3$ state to \unit[140]{s} in the $m_{s}=-3$ state and is
limited by dipolar relaxation in the former and by the finite
background gas pressure in the latter case. The optical pumping field of
about \unit[9]{G} is left on, in order to prevent thermal reoccupation
of higher $m_{s}$-levels through dipolar collision processes. During
the first \unit[5]{s} after optical pumping, we see a fast initial
decay in the atom number and a decrease in temperature, which we
ascribe to plain initial evaporation in the optical dipole trap.
To prepare a sample of up to 120\,000 atoms
at a temperature of \unit[6]{$\mu$K} and a density of
$\unit[5\cdot10^{19}]{m^{-3}}$ in a crossed optical dipole trap,
we continue the evaporation by ramping down the
intensity of the stronger of the two laser beams to \unit[5]{W}.

We look for an increase of atom loss by three-body
recombination to locate the Feshbach resonances \cite{Stenger1999c}.
This is done by first sweeping the magnetic field strength in coarse
steps on the order of \unit[0.1]{mT}--\unit[3]{mT} from \unit[0]{mT} to
\unit[60]{mT}. Smaller sweep ranges are then used in regions where
atom loss is observed. To find the precise location of the resonances a
different method is used. The magnetic field is ramped up to a value
close to the resonance in about \unit[5]{ms}.  We hold the magnetic
field for \unit[2]{s} to let the current settle and to give our
magnetic coils time to thermalize. Then the magnetic field is quickly
ramped to the desired value and held there for a variable amount of
time. The holding time is chosen to clearly resolve the resonance and
lies between \unit[100]{ms} and \unit[10]{s}.  Finally, the magnetic
field is switched off and an absorption image is taken.

The magnetic field is calibrated both slightly below and above each
resonance using RF-spectroscopy. We are able to determine the value of
the magnetic field with an one-standard deviation uncertainty of $\unit[10]{\mu T}$.

\begin{figure}[htbp!]
  \includegraphics[scale=1]{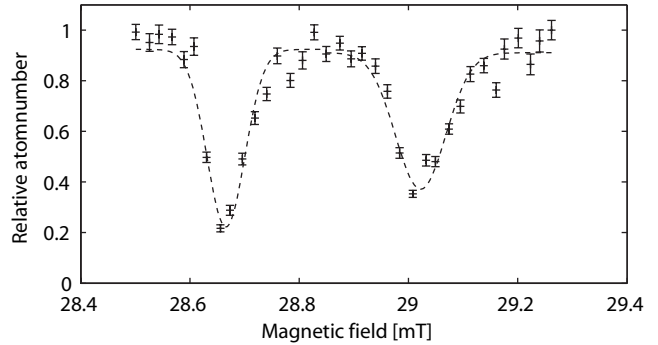}
  \caption{Inelastic loss measurement of the Feshbach resonances at
    \unit[28.66]{mT} and \unit[29.03]{mT}. The dashed lines are
    gaussian fits to the data and determine the position and width of
    the loss features.}
  \label{fig:experiment}
\end{figure}

Figure~\ref{fig:experiment} shows our data for two loss features
near \unit[29.0]{mT}. The position and widths of all the observed loss
features are determined by a gaussian fit. From the depth of these
loss features, one can estimate an upper limit for the three-body loss coefficient $L_{3}$
\cite{weber:123201}.  The error bars in the figure are obtained from
repeated measurements of atom loss and are mainly determined by number
fluctuation. All resonance parameters are tabulated in
table~\ref{tab:ResonanceData}. In addition to atom loss, we also
observe heating near most resonances, like in
\cite{weber:123201}.
The accuracy of our measurements is not limited by an
inhomogeneity in the magnetic field, as its variation across the cloud
is on the order of \unit[5]{$\mu$T}. The finite temperature of our sample
gives rise to an additional uncertainty in the resonance location that
is of the same magnitude.

\begin{figure*}[htbp!]
 \includegraphics[scale=1]{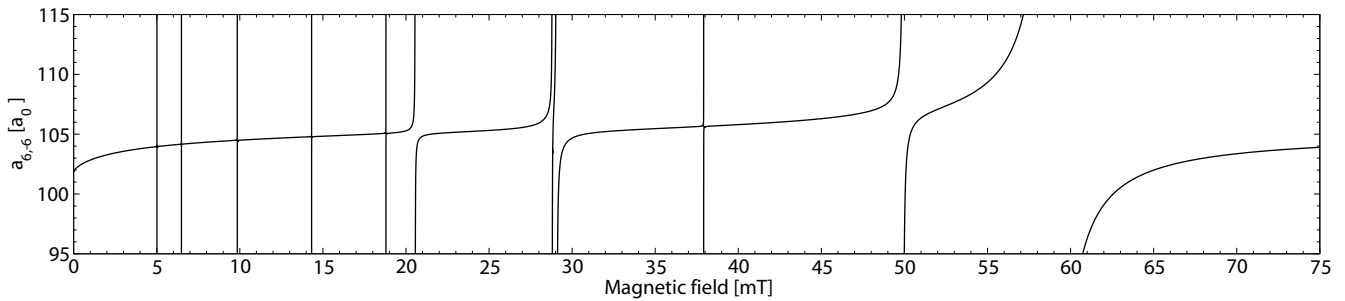}
  \caption{Calculated scattering length of two $m_s=-3$ \cro atoms versus
    magnetic field, for model parameters
    $a_6=\unit[111.56]{a_0}$, $a_4=\unit[57.61]{a_0}$,
    $a_2=-\unit[7.26]{a_0}$, $C_6=\unit[733]{a.u.}$ and $C_8=\unit[75\cdot 10^3]{a.u.}$. The feature near \unit[29]{mT} is a pair of nearly-degenerate Feshbach resonances (see Fig.~\ref{fig:experiment} for the corresponding experimental data).}
  \label{fig:theory}
\end{figure*}

Our experimental resonance positions can determine
the scattering lengths of the Born-Oppenheimer potentials
with high accuracy.
The theoretical analysis uses the Hamiltonian of a pair of $^7$S${}_{3}$ chromium atoms in an external
magnetic field $\vec B$ and includes the seven isotropic Born-Oppenheimer
potentials, the nuclear rotational energy $\hbar^2 {\vec \ell}^2 /
(2 \mu R^2)$ where $\vec \ell$
is the orbital angular momentum of the nuclei and $\mu$ the reduced mass
of the diatom, the Zeeman interaction with the magnetic field, and the
anisotropic spin-spin dipole interaction
\begin{equation}
  V_{dd}=\frac{\mu_0 (g_s \mu_B)^2}{4\pi}\frac{\vec{s}_1\cdot \vec{s}_2-3 (\vec{s}_1\cdot \hat{R})(\vec{s}_2\cdot \hat{R})}{R^3}\,.
\end{equation}
Here $\vec{s}_{1,2}$ are the electron spin of each atom,
$\vec{S}=\vec{s}_1+\vec{s}_2$, and $\hat{R}$ and $R$
are the orientation of internuclear axis and internuclear
separation, respectively. Moreover, $g_s\approx2$ is the
electron gyromagnetic ratio of \cro and $\mu_0$ is the
magnetic constant. For this Paper we do not include second-order spin-orbit
or spin-rotation interactions~\cite{Brion:1986}.

We construct $^{2S+1}\Sigma^+_{g/u}$ Born-Oppenheimer potentials $V_S$ by
smoothly joining a short-range $R\leq R_x$ model potential with the
well-known long-range dispersion potential $\sum_n -C_n/R^n$, in which we
only retain the $n=6$ and 8 terms. The connection point $R_x =
17.5~a_0$ is chosen such that each $V_S$ can be well represented by its
long-range form beyond $R_x$ and its value at $R_x$ is much larger
than the collision and bound-state energies of interest here. The
inner wall and dissociation energy of the model potentials
approximately agree with Ref.~\cite{Pavlovic:2004}. We allow for variation
of $C_n$ and include short-range corrections near the minimum
of each potential curve. This allows us to independently
tune $C_n$ and the $s$-wave scattering lengths $a_S$ of $V_S$ to fit the
experimental data.  The number of bound states of $V_S$ is uncertain
to $\pm$10 for the deeper potentials.

When the atoms are far apart, the eigenstates of the dimer are $ |
S M_S ; \ell m_{\ell} \rangle $, in which $M_{S}$ and $m_\ell$ are
projections of $\vec S$ and $\vec \ell$ along $\vec B$.  The total
projection $M=m_{\ell}+M_S$ and parity $(-1)^{\ell}$ are conserved
during the collision. As the nuclei of the atoms are identical the
selection rule $(-1)^{S + \ell}=1$ must hold. In absence of the
spin-spin interaction, the Hamiltonian conserves $\vec \ell$ and $\vec S$
as well.  The anisotropic spin-spin dipole interaction couples states with
$\Delta S = 0,2$ and $\Delta \ell = 0,2$ with $\ell=0 \to \ell^\prime=0$
transitions forbidden.

Our sample is spin polarized, so that the incoming state has quantum
numbers $S=-M_{S}=6$ by straightforward angular momentum addition.
Moreover, the temperature of the sample, $T\approx\unit[6]{\mu K}$,
is small compared to the $\ell \geq 2$ centrifugal barrier such that
incoming $\ell_i=0$ collisions dominate the scattering cross sections. We
find that, in addition to the incoming state, states with $\ell=2$ and
$4$ ($d$ and $g$) partial waves and $S=2,4$, and 6 have to be coupled
together in order to explain the $11$ strongest observed features of Table
\ref{tab:ResonanceData}. Even though, no term in the molecular Hamiltonian
directly couples $\ell=4$ states to the $\ell_i=0$ state, second order mixing
in the spin-spin dipole interaction via $\ell=2$ states is relevant in \cro.
All these states have a total projection $M=-6$.
Two of the weakest $B<1$ mT resonances in the table must be explained
with incoming $\ell_i$=2 $d$-wave collisions and $M\neq-6$.

The locations of the maxima in the
experimental three-body loss rate are compared with locations of peaks in the elastic
two-body cross section calculated by full quantum-scattering methods.
We perform a global $\chi^2$-minimization with parameters $a_{2,4,6}$,
$C_6$ and $C_8$.  Our best-fit parameters with
one standard deviation are $a_2=-\unit[7(20)]{a_0}$,
$a_4=\unit[58(6)]{a_0}$, $a_6=\unit[112(14)]{a_0}$,
$C_6=\unit[733(70)]{a.u.}$, and $C_8=\unit[75_{-75}^{+90}\cdot 10^3]{a.u.}$.
Here 1 a.u. is $E_h a_0^n$ for $C_n$ and $E_h=4.359744 \cdot 10^{-18}$ J is a Hartree.
The minimization procedure provides only a weak upper bound
on the $C_8$.
The $^{13}\Sigma_g^+$ scattering length $a_6$ is in reasonable agreement with
Ref.~\cite{Schmidt:2003b} and the $C_6$ coefficient is consistent with that of
Ref.~\cite{Pavlovic:2004}.  The average difference between theoretical
and experimental resonance positions is only $\approx
\unit[0.06]{mT}$.

Figure~\ref{fig:theory} shows the experimentally accessible $s$-wave scattering length $a_{S,M_S}$ of two colliding
$s=3$, $m_{s}=-3$ ($S=-M_S=6$) atoms as a function of magnetic field
for our best fit parameters.
Unlike the $a_S$, this scattering length depends on the spin-spin dipole interaction.
Near each Feshbach resonance, the
scattering length both diverges and crosses zero. The difference in
magnetic field between these two locations defines the resonance
width $\Delta$~\cite{Tiesinga1993a} and is given in Table~\ref{tab:ResonanceData}.

The nature of \cro Feshbach resonances can be understood through
approximate calculations of molecular bound states.
We find that calculations of eigenstates of a reduced Hamiltonian
limited to a single basis state $|SM_S;\ell m_\ell\rangle$
locates the resonances to within 0.25 mT from the scattering calculation.
Our assignment $S$, $M_S$, $\ell$, and $m_\ell$ from this approximate model
is shown in Table~\ref{tab:ResonanceData}.  An
alternative assignment in which the quantum numbers of the nearly
degenerate pair near $29.0$~mT are interchanged is
consistent with our best-fit parameters.

In the limit of vanishing spin-spin dipole interaction a simple
resonance-pattern is expected. Scattering is then independent of $m_\ell$ and
the resonances occur at $B_{\rm res}=E_B/(g_s\mu_{\rm B} (M_S+6))$,
where $E_B$ is one of the zero-field binding energies of the
potential $V_S(R)+\hbar^2\ell(\ell+1)/(2\mu R^2)$.
Inclusion of the spin-spin dipole interaction gives rise to observable
deviations from this pattern, as large as
$\approx\unit[1]{mT}$.  Such shifts are an order of magnitude
larger than the 0.06 mT discrepancies that remain after our
least-squares fit.  Moreover, the 0.06 mT agreement
strongly suggests that the spin-spin dipole
interaction is the dominant relativistic interaction in ultracold
\cro.

So far, we have focused on incoming $\ell_i=0$ $s$-wave scattering and thus
assumed $M=-6$.  We do observe resonances due to collisions from $\ell_i=2$
partial waves with  $M=-4,\dots, -8$ corresponding to
different orientations of the internuclear axis.  The pair
near $\unit[0.4]{mT}$ and $\unit[0.8]{mT}$ is due to such collisions.
These additional features are strongly suppressed at our temperatures
and we are only able to detect them at fields $B<\unit[2]{mT}$ where
we have larger atom numbers and an optimal control of the magnetic
field strength. We are not able to infer from our data a conclusive assignment of the weakest
observed resonance at \unit[0.61]{mT}.

In conclusion, we have observed Feshbach resonances
in an ultracold gas of \cro atoms held in an optical dipole trap.
Resonances were located by measuring the inelastic loss of \cro in the
energetically lowest Zeeman sublevel. Positions and widths extracted
from quantum scattering calculations are in good agreement with the
experimental data. The spin-spin dipole
interaction is essential for a quantitative understanding of the experimental
spectrum. We have improved the accuracy of the previous collisional
measurements \cite{Schmidt:2003b} and provided a determination of
the $^{9,5}\Sigma_{g}^{+}$ scattering lengths.

The resonances can be used to control the relative
strength of isotropic and anisotropic interactions. Together with the BEC of \cro we recently
realized~\cite{Griesmaier:2004}, this makes the spin-spin dipole interaction in degenerate Quantum
gases experimentally accessible.
Moreover, the formation of $\textrm{Cr}_2$ molecules via Feshbach resonances
is now possible.

\begin{table}[htbp]
  \centering
  \begin{tabular}[c]{|c|c|c|c|c|}
    \hline
    \parbox[c]{1.5cm}{Exp.\ Pos. [mT]} &
    \parbox[c]{1.5cm}{Theo.\ Pos. [mT]} &
    \parbox[c]{1.5cm}{Theo.\ $\Delta$ [$\mu$T]} &
    \parbox[c]{1.5cm}{Exp.\ L${}_{3}$ [\unitfrac{m${}^{6}$}{s}]} &
    \parbox[c]{1.8cm}{Assignment $\ell_i ;S M_S; \ell m_{\ell}$} \\ \hline
    0.41   & 0.40    & -                 &  $3\cdot 10^{-40}$ &  2; 6, -4; 0 , 0  \\ \hline
    0.61   & -       & -                 &  $8\cdot 10^{-41}$ &  -  \\ \hline
    0.82   & 0.81    & -                 &  $4\cdot 10^{-39}$ &  2; 6, -5; 0 , 0  \\\hline
    5.01   & 5.01    & $<1\cdot 10^{-4}$ &  $2\cdot 10^{-38}$ &  0; 6, -2; 4, -4\\ \hline
    6.51   & 6.49    & $6\cdot 10^{-4}$  &  $5\cdot 10^{-38}$ &  0; 6, -3; 4, -3\\ \hline
    9.89   & 9.85    & 0.030             &  $1\cdot 10^{-36}$ &  0; 6, -4; 4, -2\\ \hline
    14.39  & 14.32   & 0.012             &  $1\cdot 10^{-38}$ &  0; 4, -2; 4, -4\\ \hline
    18.83  & 18.79   & 0.022             &  $4\cdot 10^{-38}$ &  0; 4, -3; 4, -3\\ \hline
    20.58  & 20.56   & 1.2               &  $4\cdot 10^{-36}$ &  0; 6, -5; 4, -1\\ \hline
    28.66  & 28.80   & 1.2               &  $6\cdot 10^{-37}$ &  0; 4, -4; 4, -2\\ \hline
    29.03  & 29.07   & 5.1               &  $1\cdot 10^{-37}$ &  0; 6, -4; 2, -2\\ \hline
    37.92  & 37.92   & 0.042             &  $1\cdot 10^{-37}$ &  0; 2, -2; 4, -4\\ \hline
    49.99  & 49.92   & 8.1               &  $1\cdot 10^{-36}$ &  0; 4, -4; 2, -2\\ \hline
    58.91  & 58.92   & 170               &  $3\cdot 10^{-36}$ &  0; 6, -5; 2, -1\\ \hline
  \end{tabular}
  \caption{Compendium of positions and strengths of the observed
    loss features, the theoretical positions, widths, initial partial wave,
    and assignment of the resonances. Theoretical calculations use a
    collision energy of $E=k_B T$ and parameters as in Fig.~\ref{fig:theory}.
    The one standard deviation uncertainty of the experimental resonance position is $\unit[10]{\mu T}$ (See text)}
  \label{tab:ResonanceData}
\end{table}

\begin{acknowledgments}
  This work is supported within the priority programme SPP 1116 of the
  DFG and and the European RTN ``Cold Quantum Gases'' under Contract
  No. HPRN CT-2000-00125. We like to thank A. G\"orlitz and S. Giovanazzi for many
  fruitful discussions.
\end{acknowledgments}

\bibliography{crbib,bec}

\end{document}